\documentclass[twoside,version1996,preprint,9pt]{siggraph-99}
\usepackage{graphicx}
\usepackage{times}
\usepackage{fancyhdr}

\begin{document}

%----------------------------------------------------------------------------
%----------------------------------------------------------------------------

\newcommand{\BM}[1]{
  \mbox{\boldmath$#1$}
}

\newcommand{\figureTopBot}[1]{
  \begin{figure}[!tb]{\sloppy #1}\end{figure}
}

\newcommand{\figureTop}[1]{
  \begin{figure}[!t]{\sloppy #1}\end{figure}
}

\newcommand{\figureBot}[1]{
  \begin{figure}[!b]{\sloppy #1}\end{figure}
}

\newcommand{\figureWideTop}[1]{
  \begin{figure*}[!t]{\sloppy #1}\end{figure*}
}

\newcommand{\eqAlgn}{
  \!\!&\!\!
}

%----------------------------------------------------------------------------
%----------------------------------------------------------------------------

\acmcategory{research}
\acmformat{print}

 %% Who to contact about the paper
\contactname{James F. O'Brien}
\contactaddress{College of Computing\\
   801 Atlantic Drive \\
   Georgia Institute of Technology \\
   Atlanta, GA 30332-0280 }
\contactphone{(404) 894-4998}
\contactfax{(404) 894-0673}
\contactemail{obrienj@cc.gatech.edu}

 %% Keywords and estimated size of paper
\estpages{8}
\setcounter{page}{137}
%----------------------------------------------------------------------------
%----------------------------------------------------------------------------

\pagestyle{fancy}
\renewcommand{\footrulewidth}{0pt}
\thispagestyle{fancy}
\fancyfoot{}
\fancyhead{}
\fancyfoot[RO,LE]{\thepage}
\fancyhead[LE]{\fontsize{8}{8}\textsf{ACM SIGGRAPH 99, Los Angeles, California, August 8--13, 1999}}
\fancyhead[RO]{\fontsize{8}{8}\textsf{Computer Graphics Proceedings, Annual Conference Series, 1999}}

\newlength{\headrulelength}
\setlength{\headrulelength}{\textwidth}
\newlength{\headrulegap}
\setlength{\headrulegap}{.75in}
\addtolength{\headrulelength}{-\headrulegap}
\def\headrule{\hspace{\headrulegap}\rule[2ex]{\headrulelength}{\headrulewidth}\gdef\headrule{\hrule}}

\fancypagestyle{empty}{
  \fancyfoot{}
  \fancyhead{}
  \fancyhead[RO,LE]{\fontsize{8}{8}\textsf{Computer Graphics Proceedings, Annual Conference Series, 1999}}
  \fancyfoot[RO,LE]{\thepage}
  \fancyhead[LO,RE]{\includegraphics[height=.66in]{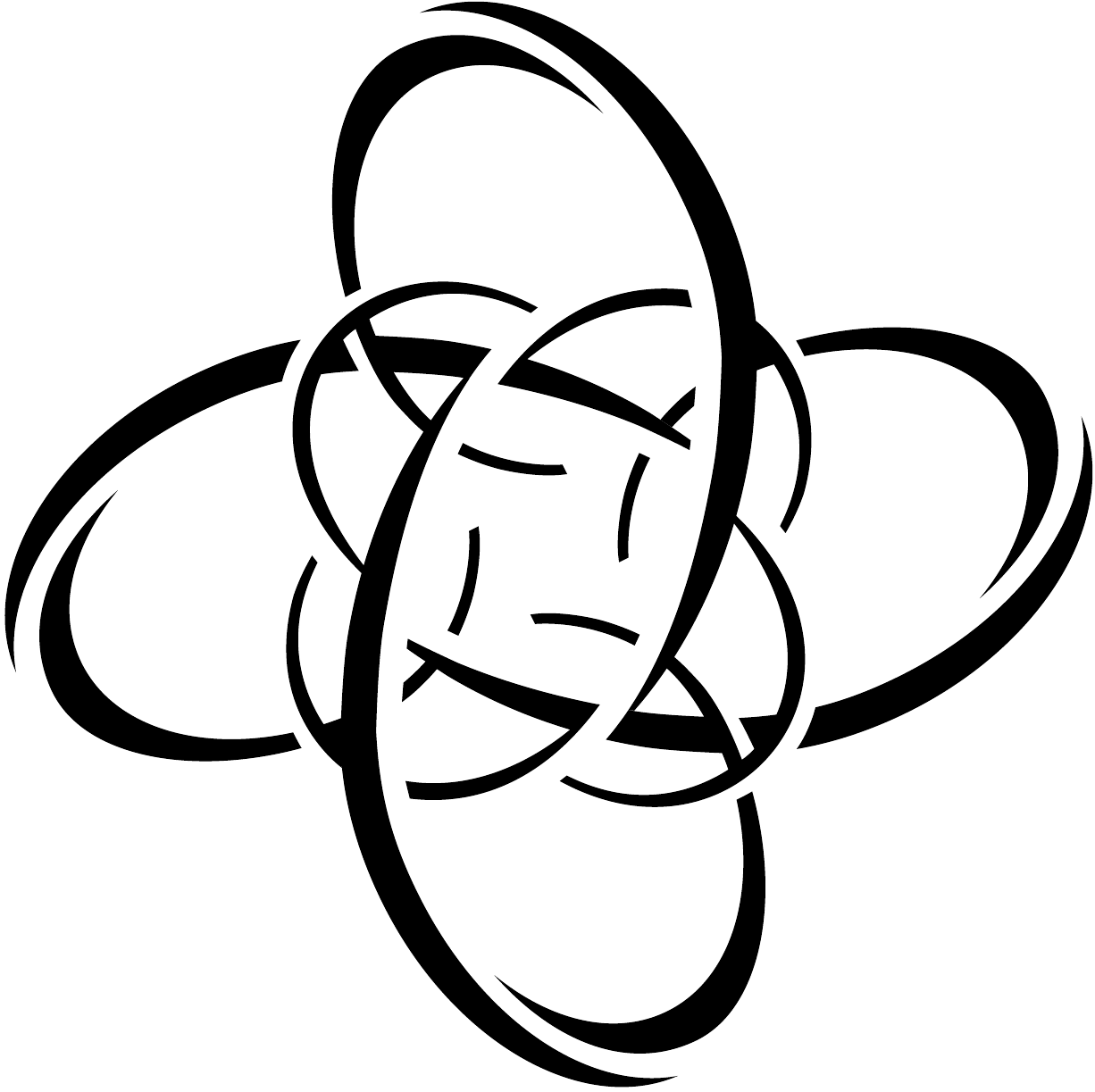}\vspace*{-.46in}}
}

%----------------------------------------------------------------------------
%----------------------------------------------------------------------------

\title{Graphical Modeling and Animation of Brittle Fracture}

\author{James F. O'Brien \and Jessica K. Hodgins}
\affiliation{
  GVU Center and College of Computing\\
  Georgia Institute of Technology
}

\maketitle

\begin{abstract}
  In this paper, we augment existing techniques for simulating
  flexible objects to include models for crack initiation and
  propagation in three-dimensional volumes.  By analyzing the stress
  tensors computed over a finite element model, the simulation
  determines where cracks should initiate and in what directions they
  should propagate.  We demonstrate our results with animations of
  breaking bowls, cracking walls, and objects that fracture when they
  collide. By varying the shape of the objects, the material
  properties, and the initial conditions of the simulations, we can
  create strikingly different effects ranging from a wall that
  shatters when it is hit by a wrecking ball to a bowl that breaks in
  two when it is dropped on edge.
\end{abstract}

\begin{CRcatlist}
  \CRcat{I.3.5}{Computer Graphics}{Computational Geometry and Object Modeling}{Physically based modeling};
  \CRcat{I.3.7}{Computer Graphics}{Three-Dimensional Graphics and Realism}{Animation};
  \CRcat{I.6.8}{Simulation and Modeling}{Types of Simulation}{Animation}
\end{CRcatlist}

\keywordlist

%----------------------------------------------------------------------------
%----------------------------------------------------------------------------

\ifcameraelse{
  \renewcommand{\thefootnote}{}%
  \footnotetext[0]{
    \par\noindent 
    College of Computing, Georgia Institute of Technology, Atlanta, GA
    30332.  job@acm.org, jkh@cc.gatech.edu.\\
    \rule[1.5in]{0in}{0in}
  } 
  \renewcommand{\thefootnote}{\arabic{footnote}}
  \preprinttext{}  
}{
  \renewcommand{\thefootnote}{}%
  \footnotetext[0]{
    \par\noindent 
    College of Computing, Georgia Institute of Technology, Atlanta, GA
    30332.  job@acm.org, jkh@cc.gatech.edu.\\
    \parbox[t][1.5in][b]{\columnwidth}{\textsf{%
      {\large \bf Authors' pre-print version.}\\
      \textbf{SIGGRAPH 99, Los Angeles, CA USA}
    }} 
  }
  \preprinttext{}  
  \renewcommand{\thefootnote}{\arabic{footnote}}
}

%----------------------------------------------------------------------------
%----------------------------------------------------------------------------

\section{Introduction}\label{sec:Introduction}

With the introduction in 1998 of simulated water in {\it
Antz}~\cite{Foster:1996:RALb,Robertson:1998:Antz} and clothing in {\it
Geri's Game}~\cite{DeRose:1998:SSC,Robertson:1998:Geri}, passive
simulation was clearly demonstrated to be a viable technique for
commercial animation.  The appeal of using simulation for objects
without an internal source of energy is not surprising, as passive
objects tend to have many degrees of freedom, making keyframing or
motion capture difficult.  Furthermore, while passive objects are
often essential to the plot of an animation and to the appearance or
mood of the piece, they are not characters with their concomitant
requirements for control over the subtle details of the motion.
Therefore, simulations in which the motion is controlled only by
initial conditions, physical equations, and material parameters are
often sufficient to produce appealing animations of passive objects.

Our approach to animating breaking objects is based on linear elastic
fracture mechanics.  We model three-dimensional volumes using a finite
element method that is based on techniques presented in the computer
graphics and mechanical engineering
literature~\cite{Cook:1989:CAF,Fung:1965:FSM,Terzopoulos:1988:DM}.  By
analyzing the stresses created as a volumetric object deforms, the
simulation determines where cracks should begin and in what directions
they should propagate.  The system accommodates arbitrary propagation
directions by dynamically re\-tesselating the mesh.  Because cracks
are not limited to element boundaries, the models can form irregularly
shaped shards and edges as they shatter.

We demonstrate the power of this approach with the following examples:
a glass slab that shatters when a weight is dropped onto
it~(Figure~\ref{fig:Slab}), an adobe wall that crumbles under the
impact of a wrecking ball~(Figure~\ref{fig:Walls}), a series of bowls
that break when they hit the floor~(Figure~\ref{fig:Bowls}), and
objects that break when they collide with each
other~(Figure~\ref{fig:TheEnd}).  To assess the realism of this
approach, we compare high-speed video images of a physical bowl
dropping onto concrete and a simulated version of the same
event~(Figure~\ref{fig:Real}).

\figureTop{
  \centerline{\includegraphics[width=\columnwidth]{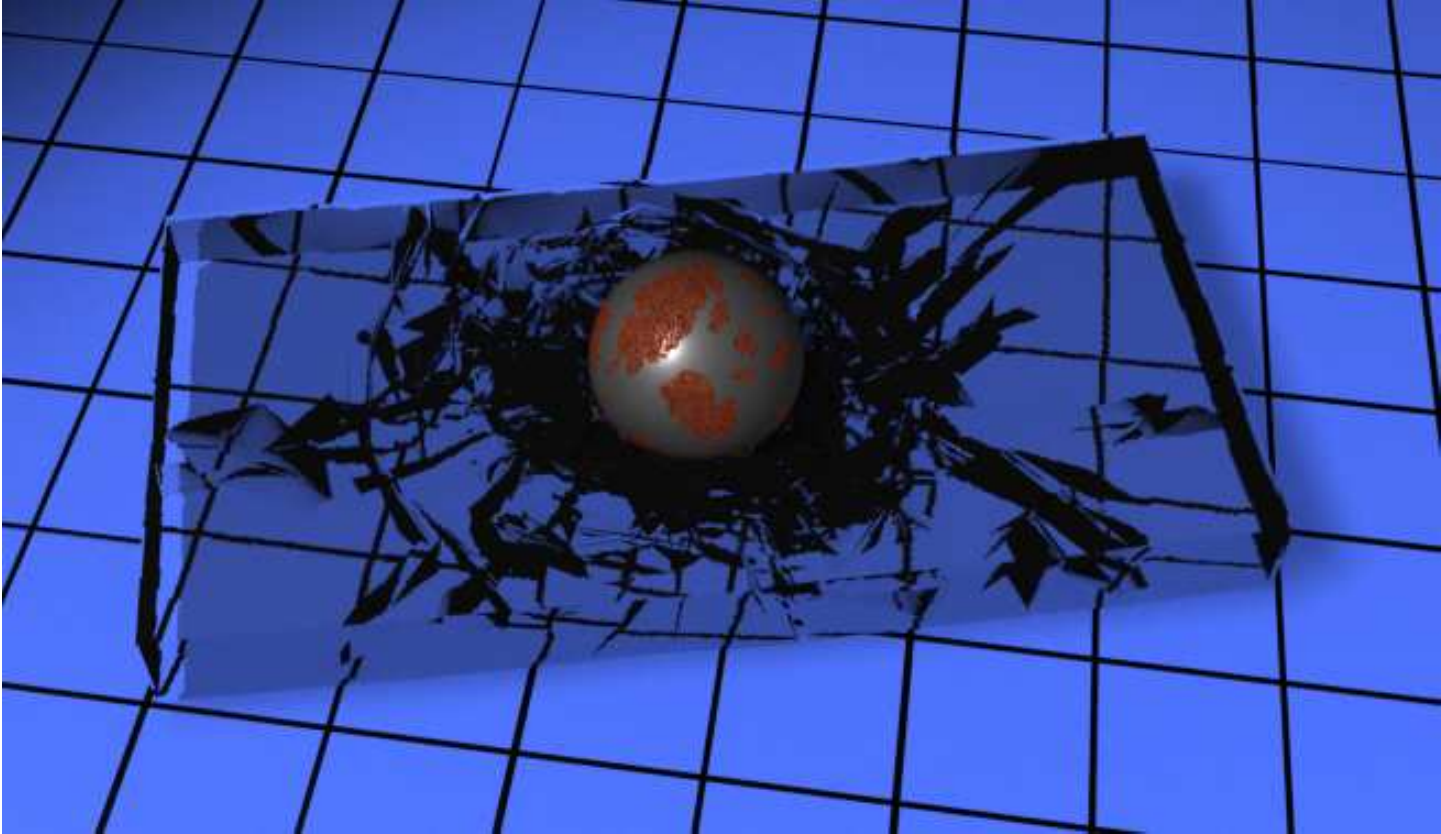}}
  \caption{ 
     Slab of simulated glass that has been shattered by a heavy weight.
  }\label{fig:Slab}
  \vspace*{-0.15in}
}

%----------------------------------------------------------------------------
%----------------------------------------------------------------------------

\section{Background}

In the computer graphics literature, two previous techniques have been
developed for modeling dynamic, deformation-induced fracture.
In~1988, Terzopoulos and
Fleischer~\cite{Terzopoulos:1988:DM,Terzopoulos:1988:MID} presented a
general technique for modeling viscoelastic and plastic deformations.
Their method used three fundamental metric tensors to define energy
functions that measured deformation over curves, surfaces, and
volumes.  These energy functions provided the basis for a continuous
deformation model that they simulated using a variety of
discretization methods.  One of their methods made use of a finite
differencing technique defined by controlled continuity
splines~\cite{Terzopoulos:1986:RIV}.  This formulation allowed them to
demonstrate how certain fracture effects could be modeled by setting
the elastic coefficients between adjacent nodes to zero whenever the
distance between the nodes exceeded a threshold.  They demonstrated
this technique with sheets paper and cloth that could be torn apart.

In~1991, Norton and his colleagues presented a technique for animating
3D solid objects that broke when subjected to large
strains~\cite{Norton:1991:AFP}.  They simulated a teapot that
shattered when dropped onto a table.  Their technique used a spring
and mass system to model the behavior of the object.  When the
distance between two attached mass points exceeded a threshold, the
simulation severed the spring connection between them.  To avoid
having flexible strings of partially connected material hanging from
the object, their simulation broke an entire cube of springs at once.

Two limitations are inherent in both of these methods.  First, when
the material fails, the exact location and orientation of the fracture
are not known.  Rather the failure is defined as the entire connection
between two nodes, and the orientation of the fracture plane is left
undefined.  As a result, these techniques can only realistically model
effects that occur on a scale much larger than the inter-node spacing.

The second limitation is that fracture surfaces are restricted to the
boundaries in the initial mesh structure.  As a result, the fracture
pattern exhibits directional artifacts, similar to the ``jaggies''
that occur when rasterizing a polygonal edge.  These artifacts are
particularly noticeable when the discretization follows a regular
pattern.  If an irregular mesh is used, then the artifacts may be
partially masked, but the fractures will still be forced onto a path
that follows the element boundaries so that the object can break apart
only along predefined facets.

Other relevant work in the computer graphics literature includes
techniques for modeling static crack patterns and fractures induced by
explosions.  Hirota and colleagues described how phenomena such as the
static crack patterns created by drying mud can be modeled using a
mass and spring system attached to an immobile
substrate~\cite{Hirota:1998:GCP}.  Mazarak~et~al.~use a voxel-based
approach to model solid objects that break apart when they encounter a
spherical blast wave~\cite{Mazarak:1999:AEO}.  Neff and Fiume use a
recursive pattern generator to divide a planar region into polygonal
shards that fly apart when acted on by a spherical blast
wave~\cite{Neff:1999:VMB}.

Fracture has been studied more extensively in the mechanics
literature, and many techniques have been developed for simulating and
analyzing the behavior of materials as they fail.  A number of
theories may be used to describe when and how a fracture will develop
or propagate, and these theories have been employed with various
numerical methods including finite element and finite difference
methods, boundary integral equations, and molecular particle
simulations.  A comprehensive review of this work can be found in the
book by Anderson~\cite{Anderson:1995:FM} and the survey article by
Nishioka~\cite{Nishioka:1997:CDF}.

Although simulation is used to model fracture both in computer
graphics and in engineering, the requirements of the two fields are
very different.  Engineering applications require that the simulation
predict real-world behaviors in an accurate and reliable fashion.  In
computer animation, what matters is how the fracture looks, how
difficult it was to make it look that way, and how long it took.
Although the technique presented in this paper was developed using
traditional engineering tools, it is an animation technique and relies
on a number of simplifications that would be unacceptable in an
engineering context.

%----------------------------------------------------------------------------
%----------------------------------------------------------------------------

\section{Deformations}\label{sec:Deformation}

Fractures arise in materials due to internal stresses created as the
material deforms.  Our goal is to model these fractures.  In order to
do so, however, we must first be able to model the deformations that
cause them.  To provide a suitable framework for modeling fractures,
the deformation method must provide information about the magnitude
and orientation of the internal stresses, and whether they are tensile
or compressive.  We would also like to avoid deformation methods in
which directional artifacts appear in the stress patterns and
propagate to the resulting fracture patterns.

We derive our deformation technique by defining a set of differential
equations that describe the aggregate behavior of the material in a
continuous fashion, and then using a finite element method to
discretize these equations for computer simulation.  This approach is
fairly standard, and many different deformation models can be derived
in this fashion.  The one presented here was designed to be simple,
fast, and suitable for fracture modeling.

%----------------------------------------------------------------------------

\subsection{Continuous Model}\label{sec:ContinuousModel}

\figureTop{
  \centerline{\includegraphics[width=\columnwidth]{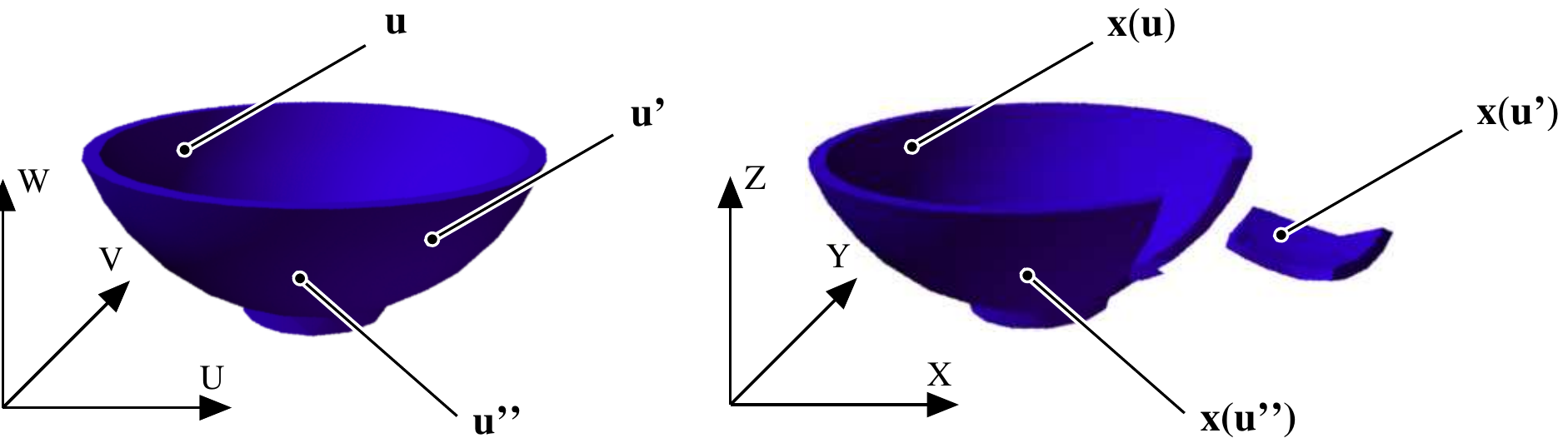}}
  \caption{ 
    The material coordinates define a 3D parameterization of the
    object.  The function $\BM{x}(\BM{u})$ maps points from their
    location in the material coordinate frame to their location in the
    world coordinates.  A fracture corresponds to a discontinuity in
    $\BM{x}(\BM{u})$.
  }\label{fig:Coords}
}

Our continuous model is based on continuum mechanics, and an excellent
introduction to this area can be found in the text by
Fung~\cite{Fung:1969:FCC}.  The primary assumption in the continuum
approach is that the scale of the effects being modeled is
significantly greater than the scale of the material's composition.
Therefore, the behavior of the molecules, grains, or particles that
compose the material can be modeled as a continuous media.  Although
this assumption is often valid for modeling deformations, macroscopic
fractures can be significantly influenced by effects that occur at
small scales where this assumption may not be valid.  Because we are
interested in graphical appearance rather than rigorous physical
correctness, we will put this issue aside and assume that a continuum
model is adequate.

We begin the description of the continuous model by defining material
coordinates that parameterize the volume of space occupied by the
object being modeled.  Let $\BM{u}=[u,v,w]^{\mathsf{T}}$ be a vector
in $\Re^3$ that denotes a location in the material coordinate frame as
shown in Figure~\ref{fig:Coords}.  The deformation of the material is
defined by the function $\BM{x}(\BM{u})=[x,y,z]^{\mathsf{T}}$ that
maps locations in the material coordinate frame to locations in world
coordinates.  In areas where material exists, $\BM{x}(\BM{u})$ is
continuous, except across a finite number of surfaces within the
volume that correspond to fractures in the material.  In areas where
there is no material, $\BM{x}(\BM{u})$ is undefined.

We make use of Green's strain tensor, $\BM{\epsilon}$, to measure the
local deformation of the material~\cite{Fung:1965:FSM}.  It can be
represented as a $3\times3$ symmetric matrix defined by
\begin{equation} \label{eq:Strain}
	\epsilon_{ij} = 
	  \left(
	    \frac{\partial \BM{x}}{\partial u_i} 
	    \cdot
	    \frac{\partial \BM{x}}{\partial u_j}
	  \right)
	  - \delta_{ij} 
\end{equation}
where $\delta_{ij}$ is the Kronecker delta: 
\begin{equation} \label{eq:Delta}
	\delta_{ij} = 
	  \left\{
	    \begin{array}{r@{\quad:\quad}l}
	      1 & i = j \\
	      0 & i \not= j                     \;.
	    \end{array}
          \right.
\end{equation}	
This strain metric only measures deformation; it is invariant with
respect to rigid body transformations applied to $\BM{x}$ and vanishes
when the material is not deformed.  It has been used extensively in
the engineering literature.  Because it is a tensor, its invariants do
not depend on the orientation of the material coordinate or world
systems.  The Euclidean metric tensor used by Terzopoulos and
Fleischer~\cite{Terzopoulos:1988:DM} differs only by the $\delta_{ij}$
term.

In addition to the strain tensor, we make use of the strain rate
tensor, $\BM{\nu}$, which measures the rate at which the strain is
changing.  It can be derived by taking the time derivative
of~(\ref{eq:Strain}) and is defined by
\begin{equation} \label{eq:StrainRate}
	\nu_{ij} =
	  \left(
	    \frac{\partial \BM{x}}{\partial u_i} 
	    \cdot
	    \frac{\partial \BM{\dot{x}}}{\partial u_j}
	  \right) +
	  \left(
	    \frac{\partial \BM{\dot{x}}}{\partial u_i} 
	    \cdot
	    \frac{\partial \BM{x}}{\partial u_j}
	  \right)
\end{equation}
where an over dot indicates a derivative with respect to time such
that $\BM{\dot{x}}$ is the material velocity expressed in world
coordinates.

The strain and strain rate tensors provide the raw information that is
required to compute internal elastic and damping forces, but they do
not take into account the properties of the material.  The stress
tensor, $\BM{\sigma}$, combines the basic information from the strain
and strain rate with the material properties and determines forces
internal to the material.  Like the strain and strain rate tensors,
the stress tensor can be represented as a $3\times3$ symmetric matrix.
It has two components: the elastic stress due to strain,
$\BM{\sigma}^{(\epsilon)}$, and the viscous stress due to strain rate,
$\BM{\sigma}^{(\nu)}$.  The total internal stress, is the sum of these
two components with
\begin{equation} \label{eq:Stress}
	\BM{\sigma} =
	  \BM{\sigma}^{(\epsilon)} +
	  \BM{\sigma}^{(\nu     )} \;.
\end{equation} 

The elastic stress and viscous stress are respectively functions of
the strain and strain rate.  In their most general linear forms, they
are defined as
\begin{equation} \label{eq:eStressGen}
	\sigma^{(\epsilon)}_{ij} = 
	  \sum_{k=1}^{3} \sum_{l=1}^{3} 
	    C_{ijkl} \, \epsilon_{kl} 
\end{equation}
\begin{equation} \label{eq:vStressGen}
	\sigma^{(\nu)}_{ij} = 
	  \sum_{k=1}^{3} \sum_{l=1}^{3} 
	    D_{ijkl} \, \nu_{kl} 
\end{equation}
where $\BM{C}$ is a set of the $81$ elastic coefficients that relate
the elements of $\BM{\epsilon}$ to the elements
$\BM{\sigma}^{(\epsilon)}$, and $\BM{D}$ is a set of the $81$ damping
coefficients.\footnote{ Actually $\BM{C}$ and $\BM{D}$ are themselves rank
four tensors, and~(\ref{eq:eStressGen}) and~(\ref{eq:vStressGen}) are
normally expressed in this form so that $\BM{C}$ and $\BM{D}$ will
follow the standard rules for coordinate transforms.}

Because both $\BM{\epsilon}$ and $\BM{\sigma}^{(\epsilon)}$ are
symmetric, many of the co\-efficients in $\BM{C}$ are either redundant
or constrained, and $\BM{C}$ can be reduced to $36$ independent values
that relate the six independent values of $\BM{\epsilon}$ to the six
independent values of $\BM{\sigma}^{(\epsilon)}$.  If we impose the
additional constraint that the material is isotropic, then $\BM{C}$
collapses further to only two independent values, $\mu$ and $\lambda$,
which are the Lam\'{e} constants of the material.
Equation~(\ref{eq:eStressGen}) then reduces to
\begin{equation} \label{eq:eStress}
	\sigma^{(\epsilon)}_{ij} = 
	  \sum_{k=1}^{3} \lambda \epsilon_{kk} \delta_{ij} +
	  2 \mu \epsilon_{ij} \;.
\end{equation}
The material's rigidity is determined by the value of $\mu$, and the
resistance to changes in volume (dilation) is controlled by 
$\lambda$.

Similarly, $\BM{D}$ can be reduced to two independent values, $\phi$
and $\psi$ and~(\ref{eq:vStressGen}) then reduces to
\begin{equation} \label{eq:vStress}
	\sigma^{(\nu)}_{ij} = 
	  \sum_{k=1}^{3} \phi \nu_{kk} \delta_{ij} +
	  2 \psi \nu_{ij}                                   \;.
\end{equation}
The parameters $\mu$ and $\lambda$ will control how quickly the
material dissipates internal kinetic energy.  Since
$\BM{\sigma}^{(\nu)}$ is derived from the rate at which ${\epsilon}$
is changing, $\BM{\sigma}^{(\nu)}$ will not damp motions that are
locally rigid, and has the desirable property of dissipating only
internal vibrations.

\figureTop{
  \centerline{\includegraphics[width=1.25in]{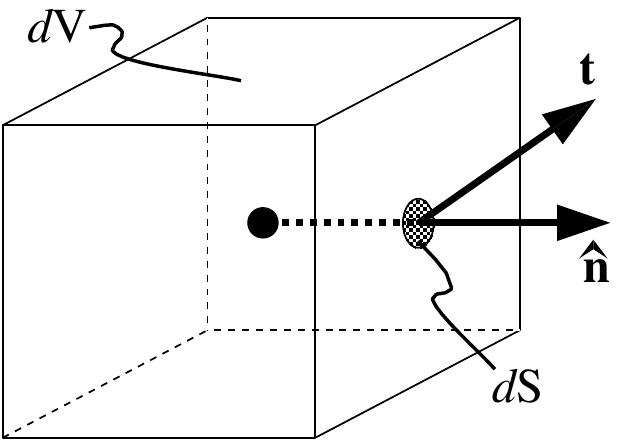}}
  \caption{ 
    Given a point in the material, the traction, $\BM{t}$, that acts
    on the surface element, $d\mathrm{S}$, of a differential volume,
    $d\mathrm{V}$, centered around the point with outward unit normal,
    $\hat{\BM{n}}$, is given by $\BM{t}=\BM{\sigma}\,\hat{\BM{n}}$.
  }\label{fig:Traction}
}

Once we have the strain, strain rate, and stress tensors, we can 
compute the elastic potential density, $\eta$, and the damping
potential density, $\kappa$, at any point in the material using
\begin{equation} \label{eq:ElasticEnergy}
	\eta = \frac{1}{2}
	  \sum_{i=1}^{3} \sum_{j=1}^{3} 
	  \sigma^{(\epsilon)}_{ij} \epsilon_{ij} \;, 
\end{equation}
%and
\begin{equation} \label{eq:DampEnergy}
	\kappa = \frac{1}{2}
	  \sum_{i=1}^{3} \sum_{j=1}^{3} 
	  \sigma^{(\nu)}_{ij} \nu_{ij} \;.
\end{equation}
These quantities can be integrated over the volume of the material to
obtain the total elastic and damping potentials.  The elastic
potential is the internal elastic energy of the material.  The damping
potential is related to the kinetic energy of the material after
subtracting any rigid body motion and normalizing for the material's
density.

The stress can also be used to compute the forces acting internal to
the material at any location.  Let $\hat{\BM{n}}$ be an outward unit
normal direction of a differential volume centered about a point in
the material. (See Figure~\ref{fig:Traction}.)  The traction (force
per unit area), $\BM{t}$, acting on a face perpendicular to the normal
is then given by
\begin{equation} \label{eq:Traction}
	\BM{t} = \BM{\sigma} \, \hat{\BM{n}} \;.
\end{equation}

The examples in this paper were generated using this isotropic
formulation.  However, these techniques do not make use of the strain
or strain rate tensors directly; rather they rely only on the stress.
Switching to the anisotropic formulation, or even to a nonlinear
stress to strain relation, would not require any significant changes.

%----------------------------------------------------------------------------

\subsection{Finite Element Discretization}

Before we can model a material's behavior using this continuous model,
it must be discretized in a way that is suitable for computer
simulation.  Two commonly used techniques are the finite difference
and finite element methods.

A finite difference method divides the domain of the material into a
regular lattice and then uses numerical differencing to approximate
the spatial derivatives required to compute the strain and strain rate
tensors.  This approach is well suited for problems with a regular
structure but becomes complicated when the structure is irregular.

A finite element method partitions the domain of the material into
distinct sub-domains, or elements as shown in Figure~\ref{fig:Mesh}.
Within each element, the material is described locally by a function
with some finite number of parameters.  The function is decomposed
into a set of orthogonal shape, or basis, functions that are each
associated with one of the nodes on the boundary of the element.
Adjacent elements will have nodes in common, so that the mesh defines
a piecewise function over the entire material domain.

\figureTop{
  \centerline{\includegraphics[width=\columnwidth]{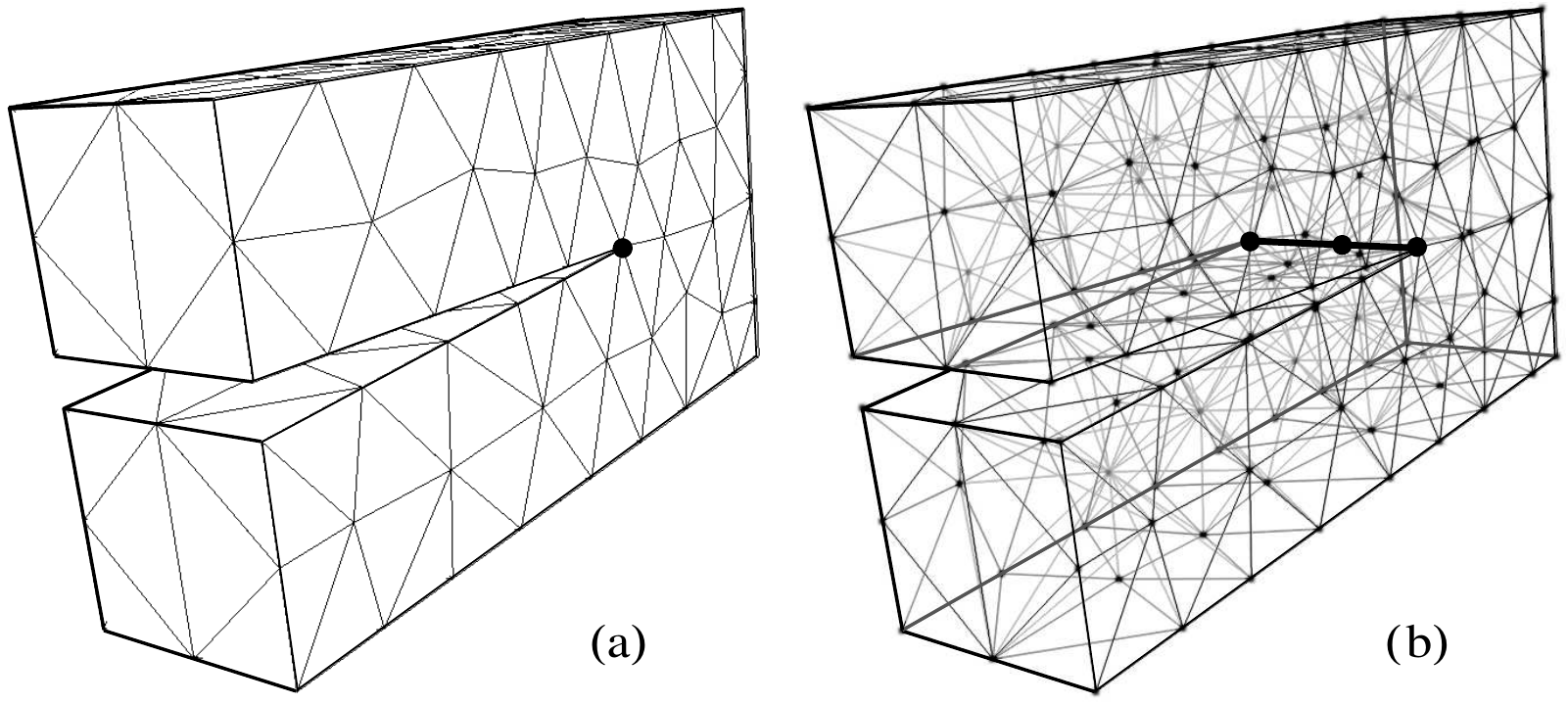}}
  \caption{ 
    Tetrahedral mesh for a simple object. In~\textbf{(a)}, only the
    external faces of the tetrahedra are drawn; in \textbf{(b)}
    the internal structure is shown.  
  }\label{fig:Mesh}
}

\figureTop{
  \centerline{\includegraphics[width=2.5in]{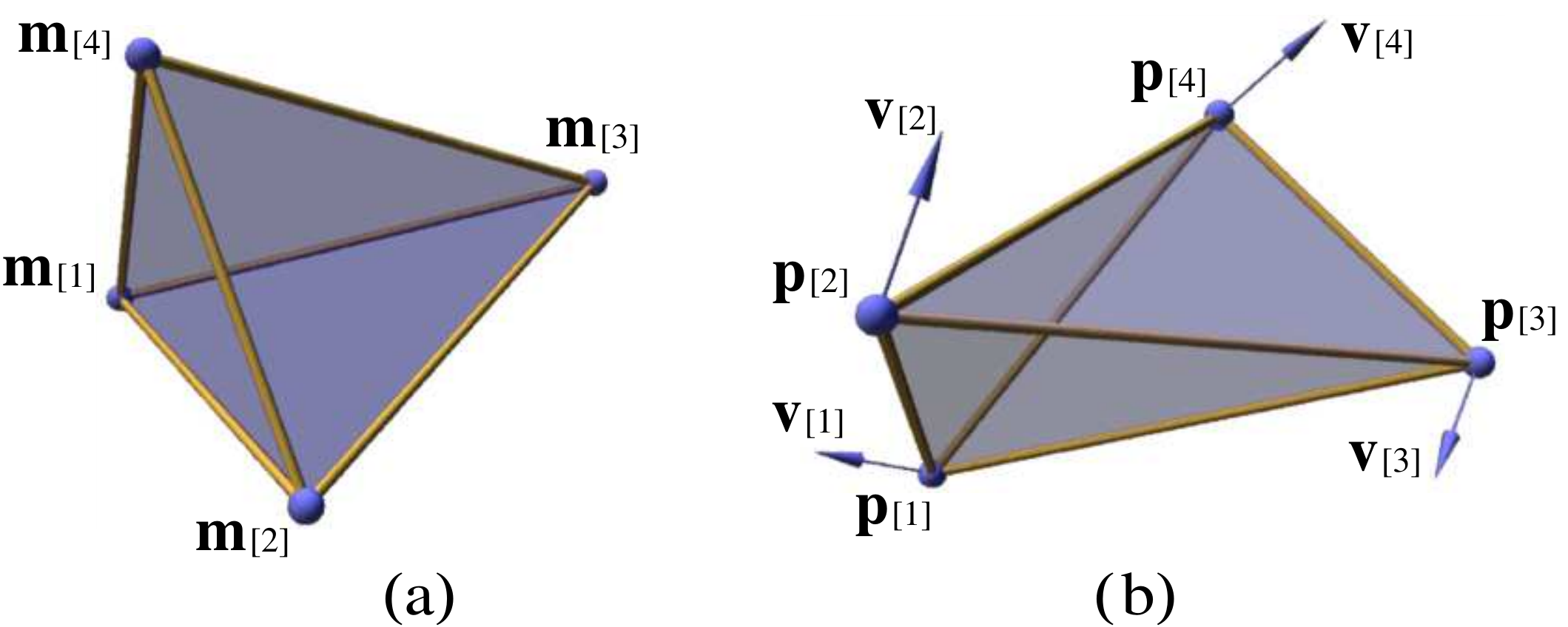}}
  \caption{ 
    A tetrahedral element is defined by its four nodes.  Each node
    has~\textbf{(a)} a location in the material coordinate system
    and~\textbf{(b)} a position and velocity in the world coordinate
    system.
  }\label{fig:Element}
}

Our discretization employs tetrahedral finite elements with linear
polynomial shape functions.  By using a finite element method, the
mesh can be locally aligned with the fracture surfaces, thus avoiding
the previously mentioned artifacts.  Just as triangles can be used to
approximate any surface, tetrahedra can be used to approximate
arbitrary volumes.  Additionally, when tetrahedra are split along a
fracture plane, the resulting pieces can be decomposed exactly into
more tetrahedra.

We chose to use linear elements because higher-order elements are not
cost effective for modeling fracture boundaries.  Although
higher-order polynomials provide individual elements with many degrees
of freedom for deformation, they have few degrees of freedom for
modeling fracture because the shape of a fracture is defined as a
boundary in material coordinates.  In contrast, with linear
tetrahedra, each degree of freedom in the world space corresponds to a
degree of freedom in the material coordinates.  Furthermore, whenever
an element is created, its basis functions must be computed.  For
high-degree polynomials, this computation is relatively expensive.
For systems where the mesh is constant, the cost is amortized over the
course of the simulation.  However, as fractures develop and parts of
the object are re\-meshed, the computation of basis matrices can
become significant.

Each tetrahedral element is defined by four nodes.  A node has a
position in the material coordinates, $\BM{m}$, a position in the
world coordinates, $\BM{p}$, and a velocity in world coordinates,
$\BM{v}$.  We will refer to the nodes of a given element by indexing
with square brackets.  For example, $\BM{m}_{[2]}$ is the position in
material coordinates of the element's second node. (See
Figure~\ref{fig:Element}.)

Barycentric coordinates provide a natural way to define the linear
shape functions within an element.  Let
$\BM{b}=[b_1,b_2,b_3,b_4]^{\mathsf{T}}$ be barycentric coordinates
defined in terms of the element's material coordinates so that
\begin{equation} \label{eq:bTerpU}
  \left[ 
    { \BM{u} \atop 1 }
  \right] 
  = 
  \left[  
    { \BM{m}_{[1]} \atop 1 } \, { \BM{m}_{[2]} \atop 1 } \,
    { \BM{m}_{[3]} \atop 1 } \, { \BM{m}_{[4]} \atop 1 }  
  \right] 
  \, \BM{b} 
  \; .
\end{equation}
These barycentric coordinates may also be used to interpolate the 
node's world position and velocity with
\begin{eqnarray} \label{eq:bTerpX}
    \left[ 
      { \BM{x} \atop 1 }
    \right] 
  \eqAlgn=\eqAlgn
    \left[  
      { \BM{p}_{[1]} \atop 1 } \, { \BM{p}_{[2]} \atop 1 } \,
      { \BM{p}_{[3]} \atop 1 } \, { \BM{p}_{[4]} \atop 1 }  
    \right] \, \BM{b} 
  \\ \label{eq:bTerpV}
    \left[ 
      { \BM{\dot{x}} \atop 1 }
    \right] 
  \eqAlgn=\eqAlgn
    \left[  
      { \BM{v}_{[1]} \atop 1 } \, { \BM{v}_{[2]} \atop 1 } \,
      { \BM{v}_{[3]} \atop 1 } \, { \BM{v}_{[4]} \atop 1 }  
    \right] \, \BM{b} 
  \;.
\end{eqnarray}

To determine the barycentric coordinates of a point within the element
specified by its material coordinates, we invert~(\ref{eq:bTerpU}) and obtain
\begin{equation} \label{eq:uTerpB}
  \BM{b} \,=\,  
    \BM{\beta} \,   
    \left[ 
      { \BM{u} \atop 1 }
    \right]  
\end{equation}
where $\BM{\beta}$ is defined by
\begin{equation} \label{eq:Beta}
  \BM{\beta} \,= 
  \left[  
    { \BM{m}_{[1]} \atop 1 } \, { \BM{m}_{[2]} \atop 1 } \,
    { \BM{m}_{[3]} \atop 1 } \, { \BM{m}_{[4]} \atop 1 }  
  \right]^{-1} \;.
\end{equation} 
Combining~(\ref{eq:uTerpB}) with~(\ref{eq:bTerpX})
and~(\ref{eq:bTerpV}) yields functions that interpolate the world
position and velocity within the element in terms of the material
coordinates:
\begin{eqnarray}
  \label{eq:uTerpX}
    \BM{     x }(\BM{u}) \eqAlgn = \eqAlgn \BM{P} \, \BM{\beta} \, \left[{\BM{u} \atop 1 }\right] \\
  \label{eq:uTerpV}
    \BM{\dot{x}}(\BM{u}) \eqAlgn = \eqAlgn \BM{V} \, \BM{\beta} \, \left[{\BM{u} \atop 1 }\right] 
\end{eqnarray}
where $\BM{P}$ and $\BM{V}$ are defined as 
\begin{eqnarray}
  \label{eq:DefnP}
    \BM{P} \eqAlgn = \eqAlgn  
      \left[  
        \BM{p}_{[1]} \, \BM{p}_{[2]} \,
        \BM{p}_{[3]} \, \BM{p}_{[4]}  
      \right] \\
  \label{eq:DefnV}
    \BM{V} \eqAlgn = \eqAlgn 
      \left[  
        \BM{v}_{[1]} \, \BM{v}_{[2]} \,
        \BM{v}_{[3]} \, \BM{v}_{[4]}  
      \right] \;.
\end{eqnarray}
Note that the rows of $\BM{\beta}$ are the coefficients of the shape
functions, and $\BM{\beta}$ needs to be computed only when an element
is created or the material coordinates of its nodes change.  For
non-degenerate elements, the matrix in~(\ref{eq:Beta}) is guaranteed
to be non-singular, however elements that are nearly co-planar will
cause $\BM{\beta}$ to be ill-conditioned and adversely affect the
numerical stability of the system.

Computing the values of $\BM{\epsilon}$ and $\BM{\nu}$ within the
element requires the first partials of $\BM{x}$ with respect to
$\BM{u}$:
\begin{eqnarray}
  \frac{\partial \BM{     x }}{\partial u_i} \eqAlgn = \eqAlgn  \BM{P} \, \BM{\beta} \, \BM{\delta}_i \\
  \frac{\partial \BM{\dot{x}}}{\partial u_i} \eqAlgn = \eqAlgn  \BM{V} \, \BM{\beta} \, \BM{\delta}_i 
\end{eqnarray}
where 
\begin{equation}
  \BM{\delta}_i = [ \delta_{i1} \, \delta_{i2} \, \delta_{i3} \, 0 ]^{\mathsf{T}} \;.
\end{equation}
Because the element's shape functions are linear, these partials are
constant within the element.

The element will exert elastic and damping forces on its nodes.  The
elastic force on the $i$th node, $\BM{f}_{[i]}^{(\epsilon)}$, is
defined as the negative partial of the elastic potential density, $\eta$, with
respect to $\BM{p}_{[i]}$ integrated over the volume of the element.
Given $\BM{\sigma}^{(\epsilon)}$, $\BM{\beta}$, and the positions in
world space of the four nodes we can compute the elastic force by
\begin{equation} \label{eq:eForce}
  \BM{f}_{[i]}^{(\epsilon)} =
	-\frac{\mathrm{vol}}{2}
	\sum_{j = 1}^{4} \BM{p}_{[j]}
	  \sum_{k = 1}^{3} 
	    \sum_{l = 1}^{3} \beta_{jl} \beta_{ik} \sigma_{kl}^{(\epsilon)}
\end{equation}
where
\begin{equation} \label{eq:Vol}
   \mathrm{vol} = \frac{1}{6}
	[ ( \BM{m}_{[2]} - \BM{m}_{[1]} ) \times 
          ( \BM{m}_{[3]} - \BM{m}_{[1]} ) ]   \cdot  ( \BM{m}_{[4]} - \BM{m}_{[1]} ) \;.
\end{equation}
Similarly, the damping force on the $i$th node,
$\BM{f}_{[i]}^{(\nu)}$, is defined as the partial of the damping
potential density, $\kappa$, with respect to $\BM{v}_{[i]}$ integrated
over the volume of the element.  This quantity can be computed with
\begin{equation} \label{eq:dForce}
  \BM{f}_{[i]}^{(\nu)} =
	-\frac{\mathrm{vol}}{2}
	\sum_{j = 1}^{4} \BM{p}_{[j]}
	  \sum_{k = 1}^{3} 
	    \sum_{l = 1}^{3} \beta_{jl} \beta_{ik} \sigma_{kl}^{(\nu)} \;.
\end{equation}
Summing these two forces, the total internal force that an element exerts
on a node is
\begin{equation} \label{eq:Force}
  \BM{f}_{[i]}^{\mathrm{el}} =
	-\frac{\mathrm{vol}}{2}
	\sum_{j = 1}^{4} \BM{p}_{[j]}
	  \sum_{k = 1}^{3} 
	    \sum_{l = 1}^{3} \beta_{jl} \beta_{ik} \sigma_{kl} \;,
\end{equation}
and the total internal force acting on the node is obtained by summing
the forces exerted by all elements that are attached to the node.

As the element is compressed to less than about $30\%$ of its material
volume, the gradient of $\eta$ and $\kappa$ start to vanish causing
the resisting forces to fall off.  We have not found this to be a
problem as even the more squishy of the materials that we have modeled
conserve their volume to within a few percent.

Using a lumped mass formulation, the mass contributed by an element to
each one of its nodes is determined by integrating the material
density, $\rho$, over the element shape function associated with that
node.  In the case of tetrahedral elements with linear shape
functions, this mass contribution is simply $\rho\,\mathrm{vol}/4$.

The derivations above are sufficient for a simulation that uses an
explicit integration scheme.  Additional work, including computing the
Jacobian of the internal forces, is necessary for implicit integration
scheme.  (See for example~\cite{Baraff:1998:LSC}
and~\cite{Cook:1989:CAF}.)

%----------------------------------------------------------------------------

\subsection{Collisions}

In addition to the forces internal to the material, the system
computes collision forces.  The collision forces are computed using a
penalty method that is applied when two elements intersect or if an
element violates another constraint such as the ground.  Although
penalty methods are often criticized for creating stiff equations, we
have found that for the materials we are modeling the internal forces
are at least as stiff as the penalty forces.  Penalty forces have the
advantage of being very fast to compute.  We have experimented with
two different penalty criteria: node penetration and overlap volume.
The examples presented in this paper were computed with the node
penetration criteria; additional examples on the conference
proceedings CD-ROM were computed with the overlap volume criteria.

The node penetration criteria sets the penalty force to be
proportional to the distance that a node has penetrated into another
element.  The penalty force acts in the direction normal to the
penetrated surface.  The reaction force is distributed over the
penetrated element's nodes so that the net force and moment are the
negation of the penalty force and moment acting at the penetrating
node.  This test will not catch all collisions, and undetected
intersecting tetrahedra may become locked together.  It is however,
fast to compute, easy to implement, and adequate for situations that
do not involve complex collision interactions.

The overlap volume criteria is more robust than the node penetration
method, but it is also slower to compute and more complex to
implement.  The intersection of two tetrahedral elements is computed
by clipping the faces of each tetrahedron against the other.  The
resulting polyhedron is then used to compute the volume and center of
mass of the intersecting region.  The area weighted normals of the
faces of the polyhedron that are contributed by one of the tetrahedra
are summed to compute the direction that the penalty force acts in.  A
similar computation can be performed for the other tetrahedra, or
equivalently the direction can be negated.  Provided that neither
tetrahedra is completely contained within the other, this criteria is
more robust than the node penetration criteria.  Additionally, the
forces computed with this method do not depend on the object
tessellation.

Computing the intersections within the mesh can be very expensive, and
we use a bounding hierarchy scheme with cached traversals to help
reduce this cost.

%----------------------------------------------------------------------------
%----------------------------------------------------------------------------

\section{Fracture Modeling}\label{sec:FractureModeling}

Our fracture model is based on the theory of linear elastic fracture
mechanics~\cite{Anderson:1995:FM}.  The primary distinction between
this and other theories of fracture is that the region of plasticity
near the crack tip\footnote{The term ``crack tip'' implies that the
fracture will have a single point at its tip.  In general, the front
of the crack will not be a single point; rather it will be a curve
that defines the boundary of the surface discontinuity in material
coordinates.  (See Figure~\ref{fig:Mesh}.)  Nevertheless, we will
refer to this front as the crack tip.}  is neglected.  Because we are
not modeling the energy dissipated by this plastic region, modeled
materials will be brittle.  This statement does not mean that they are
weak; rather the term ``brittle'' refers to the fact that once the
material has begun to fail, the fractures will have a strong tendency
to propagate across the material as they are driven by the internally
stored elastic energy.

\figureTop{
  \centerline{\includegraphics[width= .75 \columnwidth]{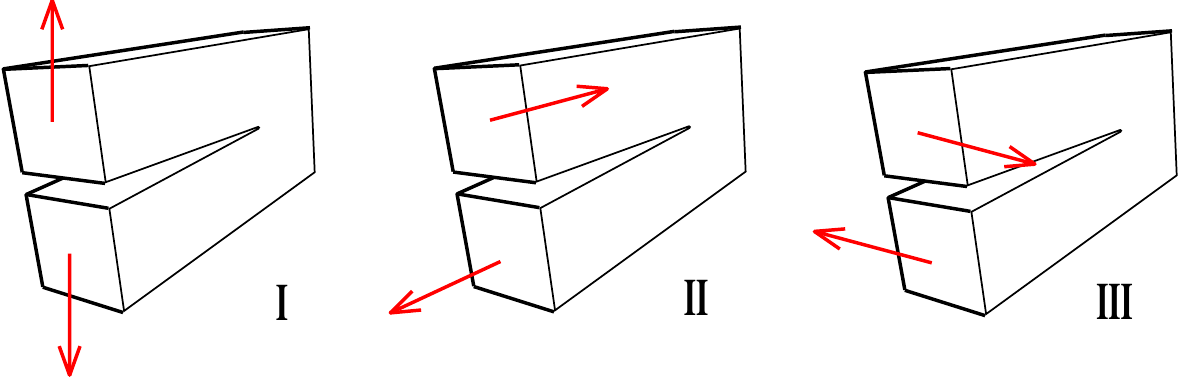}}
  \caption{ 
    Three loading modes that can be experienced by a crack.  Mode~I:
    Opening, Mode~II: In-Plane Shear, and Mode~III: Out-of-Plane
    Shear.  Adapted from Anderson~\cite{Anderson:1995:FM}.
  }\label{fig:Modes}
}

There are three loading modes by which forces can be applied to a
crack causing it to open further. (See Figure~\ref{fig:Modes}.)  In
most circumstances, some combination of these modes will be active,
producing a mixed mode load at the crack tip.  For all three cases, as
well as mixed mode situations, the behavior of the crack can be
resolved by analyzing the forces acting at the crack tip: tensile
forces that are opposed by other tensile forces will cause the crack
to continue in a direction that is perpendicular to the direction of
largest tensile load, and conversely, compressive loads will tend to
arrest a crack to which they are perpendicular.

The finite element model describes the surface of a fracture with
elements that are adjacent in material coordinates but that do not
share nodes across the internal surface.  The curve that represents
the crack tip is then implicitly defined in a piecewise linear fashion
by the nodes that border the fracture surface, and further extension
of the crack may be determined by analyzing the internal forces acting
on these nodes.

We will also use the element nodes to determine where a crack should
be initiated.  While this strategy could potentially introduce
unpleasant artifacts, we note that because the surface of an object is
defined by a polygonal boundary (the outer faces of the tetrahedra)
there will always be a node located at any concavities.  Because
concavities are precisely the locations where cracks commonly begin,
we believe that this decision is acceptable.

Our fracture algorithm is as follows: after every time step, the
system resolves the internal forces acting on all nodes into their
tensile and compressive components, discarding any un\-balanced
portions.  At each node, the resulting forces are then used to form a
tensor that describes how the internal forces are acting to separate
that node.  If the action is sufficiently large, the node is split
into two distinct nodes and a fracture plane is computed.  All
elements attached to the node are divided along the plane with the
resulting tetrahedra assigned to one or the other incarnations of the
split node, thus creating a discontinuity in the material.  Any cached
values, such as the node mass or the element shape functions, are
recomputed for the affected elements and nodes.  With this technique,
the location of a fracture or crack tip need not be explicitly
recorded unless this information is useful for some other purpose,
such as rendering.

%----------------------------------------------------------------------------

\subsection{Force Decomposition}

The forces acting on a node are decomposed by first separating the
element stress tensors into tensile and compressive components.  For a
given element in the mesh, let $\mathsf{v}^{i}(\BM{\sigma})$, with
$i\in\{1,2,3\}$, be the $i$th eigenvalue of $\BM{\sigma}$, and let
$\BM{\mathsf{\hat{n}}}^{i}(\BM{\sigma})$ be the corresponding unit
length eigenvector.  Positive eigenvalues correspond to tensile
stresses and negative ones to compressive stresses.  Since
$\BM{\sigma}$ is real and symmetric, it will have three real, not
necessarily unique, eigenvalues.  In the case where an eigenvalue has
multiplicity greater than one, the eigenvectors are selected
arbitrarily to orthogonally span the appropriate
subspace~\cite{Press:1994:NRC}.

Given a vector $\BM{a}$ in $\Re^3$, we can construct a $3\times3$
symmetric matrix, $\BM{\mathsf{m}}(\BM{a})$ that has $|\BM{a}|$ as an
eigenvalue with $\BM{a}$ as the corresponding eigenvector, and with
the other two eigenvalues equal to zero.  This matrix is defined by
\begin{equation} \label{eq:vMat}
  \BM{\mathsf{m}}(\BM{a}) =
   	  \left\{
	    \begin{array}{c@{\quad:\quad}l}
	      \BM{a} \, \BM{a}^{\mathsf{T}}/{|\BM{a}|} & \BM{a} \not= \BM{0} \\
	      \BM{0}                                   & \BM{a}     = \BM{0} \;.
	    \end{array}
          \right.
\end{equation}

The tensile component, $\BM{\sigma}^+$, and compressive component,
$\BM{\sigma}^-$, of the stress within the element can now be computed
by
\begin{eqnarray} 
  \label{eq:tSigma}
    \BM{\sigma}^{+} 
    \eqAlgn = \eqAlgn \sum_{i=1}^{3}
    \max(0,\mathsf{v}^{i}(\BM{\sigma})) \;
    \BM{\mathsf{m}}(\BM{\mathsf{\hat{n}}}^{i}(\BM{\sigma}))
  \\ \label{eq:cSigma}
    \BM{\sigma}^{-} 
    \eqAlgn = \eqAlgn \sum_{i=1}^{3}
    \min(0,\mathsf{v}^{i}(\BM{\sigma})) \;
    \BM{\mathsf{m}}(\BM{\mathsf{\hat{n}}}^{i}(\BM{\sigma}))  \;.
\end{eqnarray}

Using this decomposition, the force that an element exerts on a node
can be separated into a tensile component, $\BM{f}_{[i]}^{+}$, and a
compressive component, $\BM{f}_{[i]}^{-}$.  This separation is done by
re\-evaluating the internal forces exerted on the nodes
using~(\ref{eq:Force}) with $\BM{\sigma}^{+}$ or $\BM{\sigma}^{-}$
substituted for $\BM{\sigma}$.  Thus the tensile component is
\begin{eqnarray} 
  \label{eq:tForce}
    \BM{f}_{[i]}^{+} \eqAlgn=\eqAlgn
      -\frac{\mathrm{vol}}{2}
	\sum_{j = 1}^{4} \BM{p}_{[j]}
	  \sum_{k = 1}^{3} 
	    \sum_{l = 1}^{3} \beta_{jl} \beta_{ik} \sigma_{kl}^{+}
% \\
%  \label{eq:cForce}
%    \BM{f}_{[i]}^{-} \eqAlgn=\eqAlgn
%      -\frac{\mathrm{vol}}{2}
%	\sum_{j = 1}^{4} \BM{p}_{[j]}
%	  \sum_{k = 1}^{3} 
%	    \sum_{l = 1}^{3} \beta_{jl} \beta_{ik} \sigma_{kl}^{-}
 \;.
\end{eqnarray}
The compressive component can be computed similarly, but because
$\BM{\sigma}=\BM{\sigma}^{+}+\BM{\sigma}^{-}$, it can be computed more
efficiently using $\BM{f}_{[i]}=\BM{f}_{[i]}^{+}+\BM{f}_{[i]}^{-}$.

Each node will now have a set of tensile and a set of compressive
forces that are exerted by the elements attached to it.  For a given
node, we denote these sets as~$\{\BM{f}^{+}\}$ and~$\{\BM{f}^{-}\}$
respectively.  The unbalanced tensile load, $\BM{f}^{+}$ is simply the
sum over~$\{\BM{f}^{+}\}$, and the unbalanced compressive load,
$\BM{f}^{-}$, is the sum over~$\{\BM{f}^{-}\}$.

%----------------------------------------------------------------------------

\figureTop{
  \centerline{\includegraphics[width=\columnwidth]{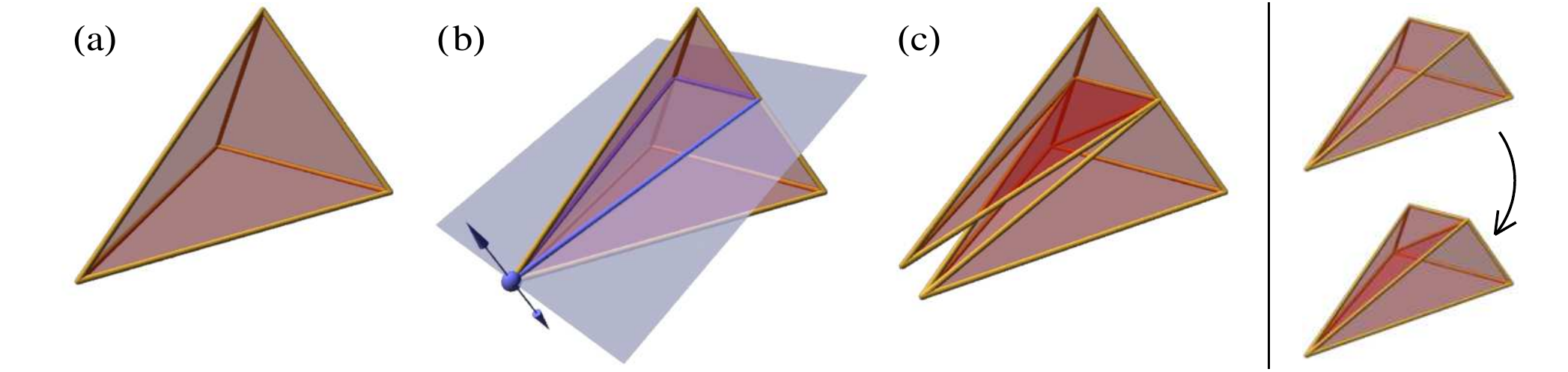}}
  \caption{ 
    Diagram showing how an element is split by the fracture plane.
    \textbf{(a)}~The initial tetrahedral element.  \textbf{(b)}~The
    splitting node and fracture plane are shown in blue.
    \textbf{(c)}~The element is split along the fracture plane into
    two polyhedra that are then decomposed into tetrahedra.  Note that
    the two nodes created from the splitting node are co-located, the
    geometric displacement shown in~\textbf{(c)} only illustrates the
    location of the fracture discontinuity.
  }\label{fig:pSplit}
  \vspace \abovecaptionskip
  \vspace \abovecaptionskip
  \centerline{\includegraphics[width=\columnwidth]{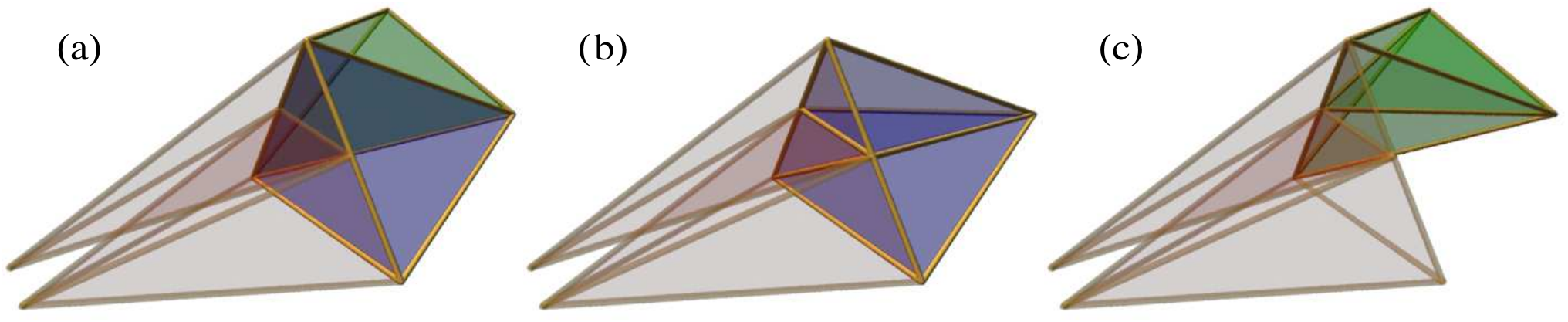}}
  \caption{ 
     Elements that are adjacent to an element that has been split by a
     fracture plane must also be split to maintain mesh consistency.
     \textbf{(a)}~Neighboring tetrahedra prior to split.
     \textbf{(b)}~Face neighbor after split.  \textbf{(c)}~Edge
     neighbor after split.
  }\label{fig:sSplit}
}

%----------------------------------------------------------------------------

\subsection{The Separation Tensor}

We describe the forces acting at the nodes using a stress variant that
we call the separation tensor, $\BM{\varsigma}$.  The separation
tensor is formed from the balanced tensile and compressive forces
acting at each node and is computed by
\begin{equation}
  \label{eq:SepTen}  
    \BM{\varsigma}  = 
      \frac{1}{2}\left(
        - \BM{\mathsf{m}}(\BM{f}^{+}) \!+\!\!\!\!\!\!
            \sum\limits_{\BM{\scriptstyle f} \in \{\BM{\scriptstyle f}^{+}\}}
              \!\!\!\!\BM{\mathsf{m}}(\BM{f}) + 
          \BM{\mathsf{m}}(\BM{f}^{-}) \!-\!\!\!\!\!\!
            \sum\limits_{\BM{\scriptstyle f} \in \{\BM{\scriptstyle f}^{-}\}}
              \!\!\!\!\BM{\mathsf{m}}(\BM{f}) 
      \right) \;.
\end{equation}
It does not respond to unbalanced actions that would produce a rigid
translation, and is invariant with respect to transformations of both
the material and world coordinate systems.

The separation tensor is used directly to determine whether a fracture
should occur at a node.  Let $\mathsf{v}^{+}$ be the largest positive
eigenvalue of $\BM{\varsigma}$.  If $\mathsf{v}^{+}$ is greater than
the material toughness, $\tau$, then the material will fail at the
node.  The orientation in world coordinates of the fracture plane is
perpendicular to $\BM{\mathsf{\hat{n}}}^{+}$, the eigenvector of
$\BM{\varsigma}$ that corresponds to $\mathsf{v}^{+}$.  In the case
where multiple eigenvalues are greater than $\tau$, multiple fracture
planes may be generated by first generating the plane for the largest
value, re\-meshing (see below), and then recomputing the new value for
$\BM{\varsigma}$ and proceeding as above.

%----------------------------------------------------------------------------

\subsection{Local Re\-meshing}

Once the simulation has determined the location and orientation of a
new fracture plane, the mesh must be modified to reflect the new
discontinuity.  It is important that the orientation of the fracture
be preserved, as approximating it with the existing element boundaries
would create undesirable artifacts. To avoid this potential
difficulty, the algorithm re\-meshes the local area surrounding the
new fracture by splitting elements that intersect the fracture plane
and modifying neighboring elements to ensure that the mesh stays
self-consistent.

First, the node where the fracture originates is replicated so that
there are now two nodes, $n^+$ and $n^-$ with the same material
position, world position, and velocity.  The masses will be
recalculated later.  The discontinuity passes ``between'' the two
co-located nodes.  The positive side of the fracture plane is
associated with $n^+$ and the negative side with $n^-$.

Next, all elements that were attached to the original node are
examined, comparing the world location of their nodes to the fracture
plane.  If an element is not intersected by the fracture plane, then
it is reassigned to either $n^+$ or $n^-$ depending on which side of
the plane it lies.  

If the element is intersected by the fracture plane, it is split along
the plane.  (See Figure~\ref{fig:pSplit}.) A new node is created along
each edge that intersects the plane.  Because all elements must be
tetrahedra, in general each intersected element will be split into
three tetrahedra.  One of the tetrahedra will be assigned to one side
of the plane and the other two to the other side.  Because the two
tetrahedra that are on the same side of the plane both share either
$n^+$ or $n^-$, the discontinuity does not pass between them.

\figureWideTop{
  \centerline{\includegraphics[width=\textwidth]{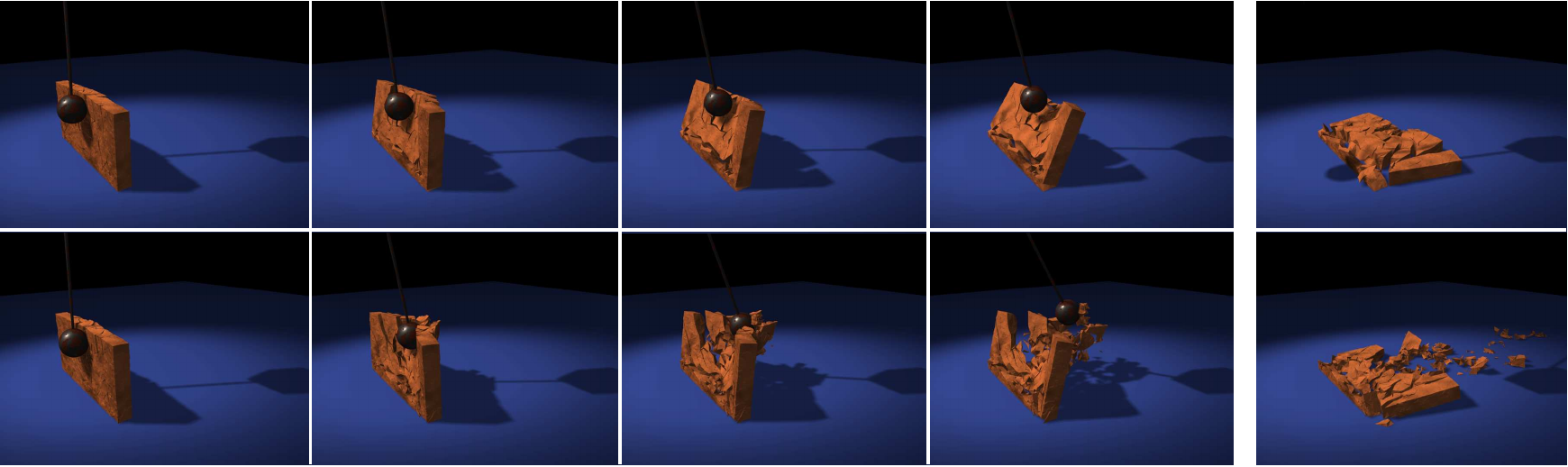}}
  \caption{ 
     Two adobe walls that are struck by wrecking balls.  Both walls
     are attached to the ground.  The ball in the second row has
     $50\times$ the mass of the first.  Images are spaced apart
     $133.3$\,ms in the first row and $66.6$\,ms in the second.  The
     rightmost images show the final configurations.
  }\label{fig:Walls}
}

\figureTop{
  \centerline{\includegraphics[width=\columnwidth]{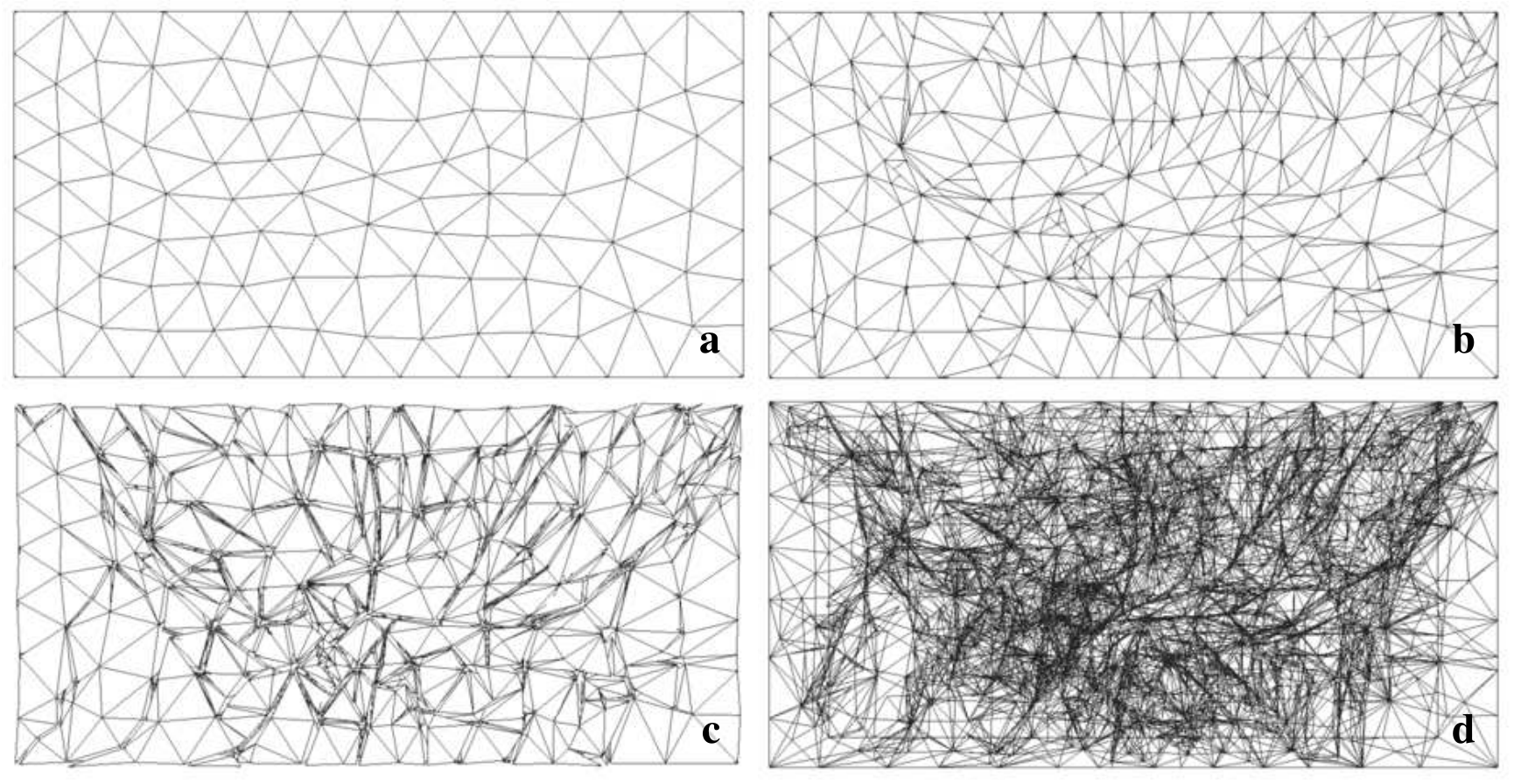}}
  \caption{ 
     Mesh for adobe wall.  \textbf{(a)}~The facing surface of the
     initial mesh used to generate the wall shown in
     Figure~\ref{fig:Walls}.  \textbf{(b)}~The mesh after being struck
     by the wrecking ball, reassembled.  \textbf{(c)}~Same as~(b),
     with the cracks emphasized. \textbf{(d)}~Internal fractures shown
     as wireframe.
  }\label{fig:WallMesh}
}

In addition to the elements that were attached to the original node,
it may be necessary to split other elements so that the mesh stays
consistent.  In particular, an element must be split if the face or
edge between it and another element that was attached to the original
node has been split.  (See Figure~\ref{fig:sSplit}.)  To prevent the
re\-meshing from cascading across the entire mesh, these splits are
done so that the new tetrahedra use only the original nodes and the
nodes created by the intersection splits.  Because no new nodes are
created, the effect of the local re\-meshing is limited to the elements
that are attached to the node where the fracture originated and their
immediate neighbors.  Because the tetrahedra formed by the secondary
splits do not attach to either $n^+$ or $n^-$, the discontinuity does
not pass between them.

Finally, after the local re\-meshing has been completed, any cached
values that have become invalid must be recomputed.  In our
implementation, these values include the element basis matrix,
$\BM{\beta}$, and the node masses. 

Two additional subtleties must also be considered.  The first subtlety
occurs when an intersection split involves an edge that is formed only
by tetrahedra attached to the node where the crack originated.  When
this happens, the fracture has reached a boundary in the material, and
the discontinuity should pass through the edge.  Re\-meshing occurs as
above, except that two nodes are created on the edge and one is
assigned to each side of the discontinuity.

Second, the fracture plane may pass arbitrarily close to an existing
node producing arbitrarily ill-conditioned tetrahedra. To avoid this,
we employ two thresholds, one the distance between the fracture plane
and an existing node, and the other on the angle between the fracture
plane and a line from the node where the split originated to the
existing node.  If either of these thresholds are not met, then the
intersection split is snapped to the existing node.  In our results,
we have used thresholds of $5$\,mm and $0.1$\,{radians}.

%----------------------------------------------------------------------------
%----------------------------------------------------------------------------

\section{Results and Discussion}

\figureWideTop{
  \centerline{\includegraphics[width=\textwidth]{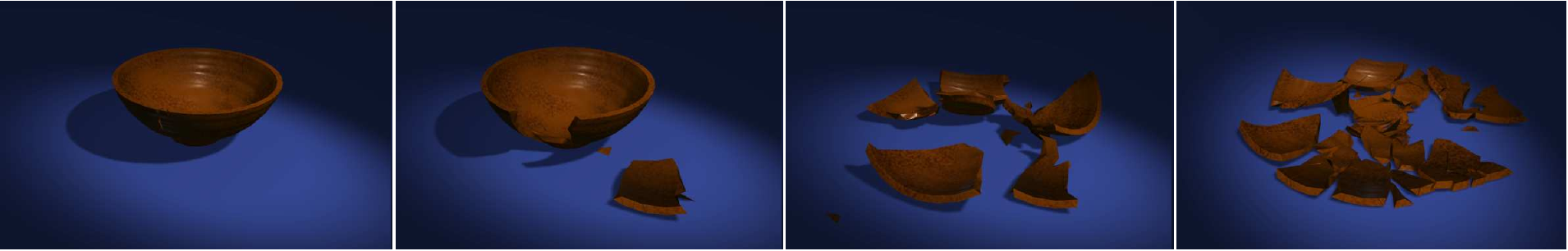}}
  \caption{ 
     Bowls with successively lower toughness values, $\tau$.  Each of
     the bowls were dropped from the same height.  Other than $\tau$,
     the bowls have same material properties.
  }\label{fig:Bowls}
}

To demonstrate some of the effects that can be generated with this
fracture technique, we have animated a number of scenes that involve
objects breaking.  Figure~\ref{fig:Slab} shows a plate of glass that
has had a heavy weight dropped on it.  The area in the immediate
vicinity of the impact has been crushed into many small fragments.
Further away from the weight, a pattern of radial cracks has
developed. 

Figure~\ref{fig:Walls} shows two walls being struck by wrecking balls.
In the first sequence, the wall develops a network of cracks as it
absorbs most of the ball's energy during the initial impact.  In the
second sequence, the ball's mass is $50\times$ greater, and the wall
shatters when it is struck.  The mesh used to generate the wall
sequences is shown in Figure~\ref{fig:WallMesh}.  The initial mesh
contains only $338$~nodes and $1109$~elements.  By the end of the
sequence, the mesh has grown to $6892$~nodes and $8275$~elements.
These additional nodes and elements are created where fractures occur;
a uniform mesh would require many times this number of nodes and
elements to achieve a similar result.

Figure~\ref{fig:Bowls} shows the final frames from four animations of
bowls that were dropped onto a hard surface.  Other than the
toughness, $\tau$, of the material, the four simulations are
identical.  The first bowl develops only a few cracks; the weakest
breaks into many pieces.

Because this system works with solid tetrahedral volumes rather than
with the polygonal boundary representations created by most modeling
packages, boundary models must be converted before they can be used.
A number of systems are available for creating tetrahedral meshes from
polygonal boundaries.  The models that we used in these examples were
generated either from a CSG description or a polygonal boundary
representation using NETGEN, a publicly available mesh generation
package~\cite{Netgen}.

Although our approach avoids the ``jaggy'' artifacts in the fracture
patterns caused by the underlying mesh, there remain ways in which the
results of a simulation are influenced by the mesh structure.  The
most obvious is that the deformation of the material is limited by the
degrees of freedom in the mesh, which in turn limits how the material
can fracture.  This limitation will occur with any discrete system.
The technique also limits where a fracture may initiate by examining
only the existing nodes.  This assumption means that very coarse mesh
sizes might behave in an unintuitive fashion.  However, nodes
correspond to the locations where a fracture is most likely to begin;
therefore, with a reasonable grid size, this limitation is not a
serious handicap.

A more serious limitation is related to the speed at which a crack
propagates.  Currently, the distance that a fracture may travel during
a time step is determined by the size of the existing mesh elements.
The crack may either split an element or not; it cannot travel only a
fraction of the distance across an element.  If a crack were being
opened slowly by an applied load on a model with a coarse resolution
mesh, this limitation would lead to a ``button popping'' effect where
the crack would travel across one element, pause until the stress
built up again, and then move across the next element.  A second type
of artifact may occur if the crack's speed should be significantly
greater than the element width divided by the simulation time step.
In this case, a high stress area will race ahead of the crack tip,
causing spontaneous failures to occur in the material.  Although we
have not observed these phenomena in our examples, developing an
algorithm that allows a fracture to propagate arbitrary distances is
an area for future work.

\figureTopBot{
  \centerline{\includegraphics[width=\columnwidth]{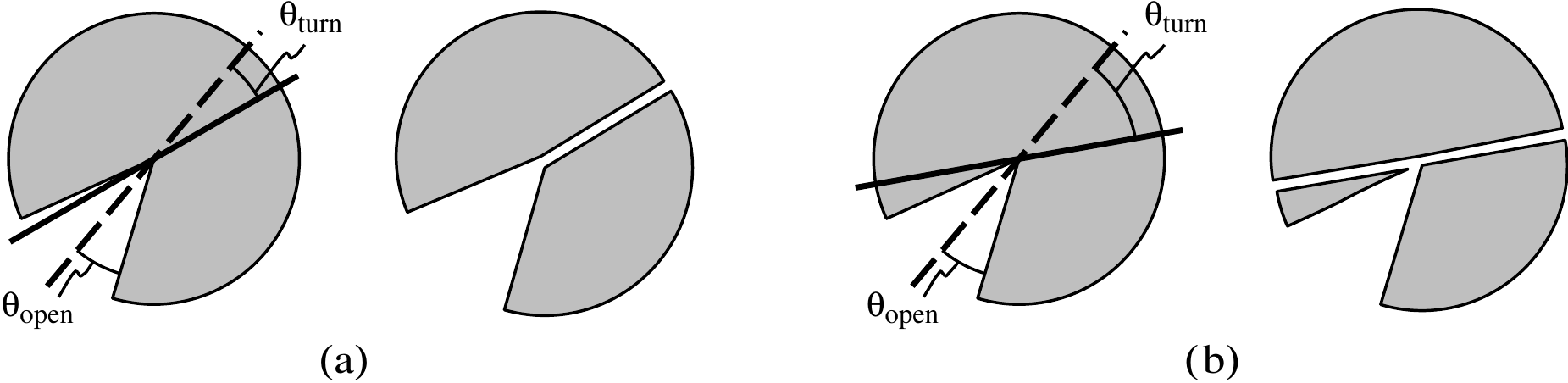}}
  \caption{
    Back-cracking during fracture advance. The dashed line is the axis
    of the existing crack.  Cracks advance by splitting elements along
    a fracture plane, shown as a solid line, computed from the
    separation tensor.  \textbf{(a)}~If the crack does not turn
    sharply, then only elements in front of the tip will be split.
    \textbf{(b)}~If the crack turns at too sharp an angle, then the
    backwards direction may not fall inside of the existing failure
    and a spurious bifurcation will occur.
  }\label{fig:BackCrack}
}

Another limitation stems from the fact that while the fracture plane's
orientation is well defined, the crack tip's forward direction is not.
As shown in Figure~\ref{fig:BackCrack}, if the cracks turns at an
angle greater than half the angle at the crack tip, then a secondary
fracture will develop in the opposite direction to the crack's
advance.  While this effect is likely present in some of our examples,
it does not appear to have a significant impact on the quality of the
results.  If the artifacts were to be a problem, they could be
suppressed by explicitly tracking the fracture propagation directions
within the mesh.

\newcommand{\C}[1]{\multicolumn{1}{c}{#1}}
\newcommand{\Cb}[1]{\multicolumn{1}{c|}{#1}}
\newcommand{\Cbb}[1]{\multicolumn{1}{c||}{#1}}

\newcommand{\R}{\rule{0ex}{2.5ex}}

\begin{table*}[tb]
  \centerline{\small
    \begin{tabular}{lc||cc|cc|cc||c|c|c}
      \hline
      \hline
                  &                 & \multicolumn{6}{c||}{                   } & \multicolumn{3}{c}{Minutes of Computation}\\    
                  &                 & \multicolumn{6}{c||}{Material Parameters} & \multicolumn{3}{c}{Time per Simulation Second}\\
      Example     &Figure           & \C{$\lambda$ ($N/m^2$)} &  \Cb{$\mu$ ($N/m^2$)} & \C{$\phi$ ($Ns/m^2$)} & \Cb{$\psi$ ($Ns/m^2$)} & \C{$\rho$ ($kg/m^3$)} & \Cbb{$\tau$ ($N/m^2$)} & \Cb{~Minimum~} & \Cb{~Maximum~} & \C{~Average~} \\
      \hline
      \hline
      Glass     \R&\ref{fig:Slab}   & $1.04 \times 10^8$    &  $1.04 \times 10^8$    & $0$         &   $6760$     &  $2588$    & $10140$   &   $75$   &  $667$   & $273$   \\  % SF = 337.947244
      Wall \#1  \R&\ref{fig:Walls}.a& $6.03 \times 10^8$    &  $1.21 \times 10^8$    & $3015$      &   $6030$     &  $2309$    & $6030$    &   $75$   &  $562$   & $399$   \\  % SF = 301.5
      Wall \#2  \R&\ref{fig:Walls}.b& $0$                   &  $1.81 \times 10^8$    & $0$         &   $18090$    &  $2309$    & $6030$    &   $75$   &  $2317$  & $1098$  \\  % SF = 301.5
      \hline
      Bowl \#1  \R&\ref{fig:Bowls}.a& $2.65 \times 10^6$    &  $3.97 \times 10^6$    & $264$       &   $397$      &  $1013$    & $52.9$    &   $90$   &  $120$   & $109$   \\  % SF = 132.25 %Sim name is Bowl-06
      Bowl \#2  \R&\ref{fig:Bowls}.b& $2.65 \times 10^6$    &  $3.97 \times 10^6$    & $264$       &   $397$      &  $1013$    & $39.6$    &   $82$   &  $135$   & $115$   \\  % SF = 132.25 %Sim name is Bowl-03
      Bowl \#3  \R&\ref{fig:Bowls}.c& $2.65 \times 10^6$    &  $3.97 \times 10^6$    & $264$       &   $397$      &  $1013$    & $33.1$    &   $90$   &  $150$   & $127$   \\  % SF = 132.25 %Sim name is Bowl-04
      Bowl \#4  \R&\ref{fig:Bowls}.d& $2.65 \times 10^6$    &  $3.97 \times 10^6$    & $264$       &   $397$      &  $1013$    & $13.2$    &   $82$   &  $187$   & $156$   \\  % SF = 132.25 %Sim name is Bowl-02
      \hline
      Comp. Bowl\R&\ref{fig:Real}   & $0$                   &  $5.29 \times 10^7$    & $0$         &   $198$      &  $1013$    & $106$     &   $247$  &  $390$   & $347$   \\  % SF = 132.25 %Sim name is Bowl-07-4
      The End   \R&\ref{fig:TheEnd} & $0$                   &  $9.21 \times 10^6$    & $0$         &   $9.2$      &  $705$     & $73.6$    &   $622$  &  $6667$  & $4665$  \\  % SF = 92.060590
      \hline
      \hline
    \end{tabular}
  }
  \caption{ 
    Material parameters and simulation times for examples.  The times
    listed reflect the total number of minutes required to compute one
    second of simulated data, including graphics and file I/O.  Times
    were measured on an SGI O2 with a $195$\,MHz MIPS R10K processor.
  }\label{tab:PnT}
\end{table*}

The simulation parameters used to generate the examples in this paper
are listed in Table~\ref{tab:PnT} along with the computation time
required to generate one second of animation.  While the material
density values, $\rho$, are appropriate for glass, stone, or ceramic,
we used values for the Lam\'{e} constants, $\lambda$ and $\mu$, that
are significantly less than those of real materials.  Larger values
would make the simulated materials appear stiffer, but would also
require smaller time steps.  The values that we have selected
represent a compromise between realistic behavior and reasonable
computation time.

Our current implementation can switch between either a forward Euler
integration scheme or a second order Taylor integrator.  Both of these
techniques are explicit integration schemes, and subject to stability
limits that require very small time steps for stiff materials.
Although semi-implicit integration methods have error bounds similar
to those of explicit methods, the semi-implicit integrators tend to
drive errors towards zero rather than infinity so that they are stable
at much larger time steps.  Other researchers have shown that by
taking advantage of this property, a semi-implicit integrator can be
used to realize speed ups of two or three orders of magnitude when
modeling object deformation~\cite{Baraff:1998:LSC}.  Unfortunately, it
may be difficult to realize these same improvements when fracture
propagation is part of the simulation.  As discussed above, the crack
speed is limited in inverse proportion to the time step size, and the
large time steps that might be afforded by a semi-implicit integrator
could cause spontaneous material failure to proceed crack advance.  We
are currently investigating how our methods may be modified to be
compatible with large time step integration schemes.

Many materials and objects in the real world are not homogeneous, and
it would be interesting to develop graphical models for animating them
as they fail.  For example, a brick wall is made up of mortar and
bricks arranged in a regular fashion, and if simulated in a situation
like our wall example, a distinct pattern would be created.
Similarly, the connection between a handmade cup and its handle is
often weak because of the way in which the handle is attached.

One way to assess the realism of an animation technique is by
comparing it with the real world.  Figure~\ref{fig:Real} shows
high-speed video footage of a physical bowl as it falls onto its edge
compared to our imitation of the real-world scene.  Although the two
sets of fracture patterns are clearly different, the simulated bowl
has some qualitative similarities to the real one.  Both initially
fail along the leading edge where they strike the ground, and
subsequently develop vertical cracks before breaking into several
large pieces.

\figureWideTop{
  \centerline{\includegraphics[width=\textwidth]{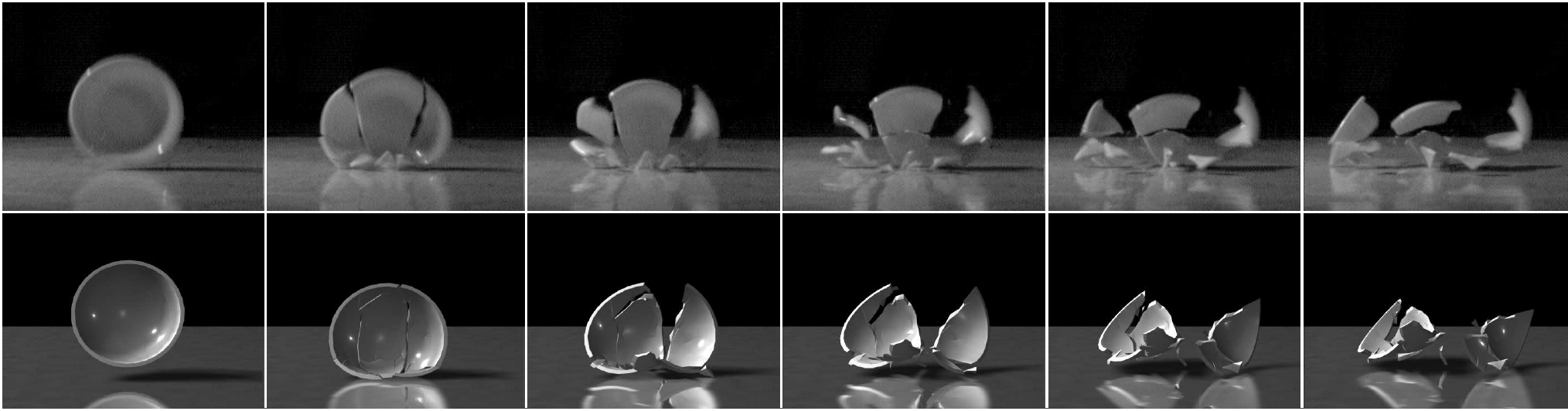}}
  \caption{ 
     Comparison of a real-world event and simulation.  The top row
     shows high-speed video images of a physical ceramic bowl dropped
     from approximately one meter onto a hard surface.  The bottom row
     is the output from a simulation where we attempted to match the
     initial conditions of the physical bowl.  Video images are
     $8$\,ms apart. Simulation images are $13$\,ms apart.
  }\label{fig:Real}
}

%In addition to colliding breakable objects with rigid shapes, we can
%collide them with each other as shown in Figure~\ref{fig:Table}.
%
%\figureWideTop{
%  \centerline{\includegraphics[width=\textwidth]{PdfFigures/table.pdf}}
%  \caption{
%    Two breakable objects colliding with each other.  Both the table
%    and the object falling on it are modeled using our fracture
%    technique.
%  }\label{fig:Table}
%}

%----------------------------------------------------------------------------
%----------------------------------------------------------------------------

\section*{Acknowledgments}

The authors would like to thank Wayne L. Wooten of Pixar Animation
Studios for lighting, shading, and rendering the images for many of
the figures in this paper.  We would also like to thank Ari Glezer and
Bojan Vukasinovic of the School Mechanical Engineering at the Georgia
Institute of Technology for their assistance and the use of the
high-speed video equipment.  Finally, we would like to thank those in
the Animation Lab who lent a hand to ensure that we made the
submission deadline.

This project was supported in part by NSF NYI Grant No.~IRI-9457621,
Mitsubishi Electric Research Laboratory, and a Packard Fellowship.
The first author was supported by a Fellowship from the Intel
Foundation.

%----------------------------------------------------------------------------
%----------------------------------------------------------------------------

\bibliographystyle{plain}

\figureBot{
  \centerline{\includegraphics[width=\columnwidth]{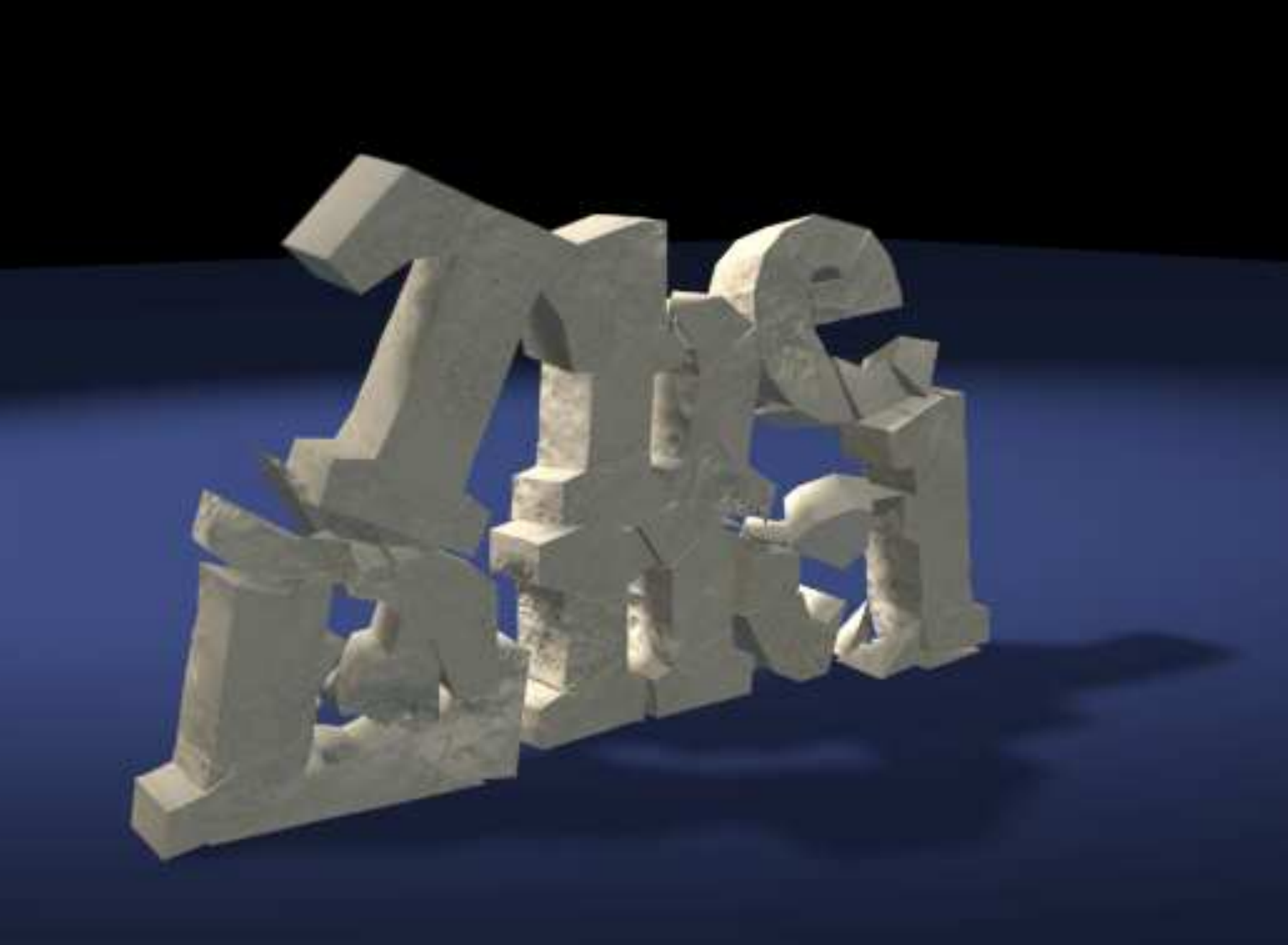}}
  \caption{ 
    Several breakable objects that were dropped from a height.
  }\label{fig:TheEnd}
}

\bibliography{fracture}

%----------------------------------------------------------------------------
%----------------------------------------------------------------------------

%\clearpage \onecolumn
%\vspace*{-0.3in}
%  \centerline{
%    \includegraphics[width=3.33in]{PdfFigures/slab.pdf}
%    \hskip 0.3in
%    \parbox[b]{3.33in}{
%      \includegraphics[width=3.33in]{PdfFigures/sSplit.pdf}
%    \vskip 0.2in
%      \includegraphics[width=3.33in]{PdfFigures/coords.pdf}
%    }
%  }
%\vskip 0.2in
%  \centerline{
%    \includegraphics[width=3.33in]{PdfFigures/pSplit.pdf}
%    \hskip 0.3in
%    \includegraphics[width=2.5in]{PdfFigures/elementNew.pdf}
%  }
%\vskip 0.2in
%  \centerline{\includegraphics[width=7in]{PdfFigures/walls.pdf}}
%\vskip 0.2in
%  \centerline{\includegraphics[width=7in]{PdfFigures/bowls.pdf}}
%\vskip 0.2in
%  \centerline{\includegraphics[width=7in]{PdfFigures/real.pdf}}
%\clearpage

%----------------------------------------------------------------------------
%----------------------------------------------------------------------------

\end{document}
\endinput

%----------------------------------------------------------------------------
%----------------------------------------------------------------------------